\documentclass[nofootinbib,showpacs,preprintnumbers,amsmath,amssymb,floatfix]
{revtex4}
\usepackage{graphicx}
\usepackage{epsf}
\usepackage{wrapfig}

\begin{document}


\title{Light-by-light scattering in Double-Logarithmic Approximation}

\vspace*{0.3 cm}

\author{B.I.~Ermolaev}
\affiliation{Ioffe Physico-Technical Institute, 194021 St.Petersburg, Russia}

\author{S.I.~Troyan}
\affiliation{St.Petersburg Institute of Nuclear Physics, 188300
Gatchina, Russia}

\begin{abstract}
In the present
paper we
consider the elastic $2 \to 2$ -scattering of virtual photons at high energies in the forward kinematics at zero and non-zero values of $t$.
Accounting for both gluon and quark double-logarithmic (DL) contributions to all orders in the QCD coupling, we
obtain explicit expressions for amplitudes  of this process in Double-Logarithmic Approximation (DLA).
First we keep the QCD coupling fixed and then account for running coupling effects.
Applying the saddle-point method to the obtained expressions for the scattering amplitude, we calculate the
high-energy asymptotics of
the amplitude, which proved to be of the Regge form. The Reggeon bears the vacuum quantum
numbers and therefore it is a new, DL contribution to Pomeron. Comparison of the DL Pomeron to the BFKL Pomeron
shows that contribution of the DL Pomeron to the high-energy asymptotics is of the same order as
contribution of the BFKL Pomeron, so the DL Pomeron should be taken into account together with the BFKL Pomeron.
We estimate the applicability region for the asymptotics of the light-by-light scattering amplitude, where the
the DL Pomeron can reliably represent the parent amplitude.
\end{abstract}


\maketitle

\section{Introduction}

Elastic scattering of virtual unpolarized photons

\begin{equation}\label{gamma}
\gamma^* (p)~ \gamma^* (q) \to  \gamma^* (p')~ \gamma^*(q')
\end{equation}

at high energies in the forward kinematics

\begin{equation}\label{fkin}
s = (p + q)^2  \sim - u =  -(p-q')^2  \gg - t = - (p - p')^2,  Q^2_1, Q^2_2,
\end{equation}
where $Q^2_1 = |p^2|  \approx |p'^2|$ and $Q^2_2 = |q^2| \approx |q'^2|$,
has been an interesting object for theoretical investigation.
The point is that theoretical investigation of
hadronic
reactions involves parton distributions. They inevitably involve non-perturbative phenomenological
contributions. In contrast,
the process (\ref{gamma}) can be studied with the means of Perturbative QCD entirely.  Because of that
remarkable feature, the light-by-light scattering was the first physics process to apply
both BFKL\cite{lobfkl,nlobfkl}.
and the approaches~\cite{photonsBFKL}- \cite{Ivanov2014hpa} based on development of BFKL.
Another attraction to study this process is that this process in
studied experimentally by ATLAS Collaboration\cite{atlas}.

The external photons in the process (\ref{gamma})
can interact through quark loops only. First of all, it can be just a single quark loop
without QCD radiative corrections. Throughout the paper we address this case as the "Born" approximation.
This approximation was applied to
analysis of the ATLAS results in Refs.~\cite{2,3,4}.

\begin{figure}[h]
  \includegraphics[width=.2\textwidth]{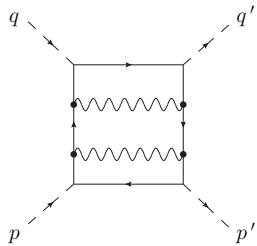}
  \caption{\label{gammaladfig1}The dashed lines denote the external photons, the straight lines form the
  single quark
loop and the wavy lines denote virtual gluons.}
\end{figure}

The Born case can be complemented by insertion of gluon propagators into the same single quark loop. An example of
the involved Feynman graphs is shown in Fig.~1. The leading contributions in this case are double-logarithmic (DL).
In the simplest version, series of DL contributions to a scattering amplitude $A$ looks as follows:

\begin{equation}\label{dlser}
A_{DL} = A_{Born} \left[1 + c^{DL}_1 \alpha_s \ln^2 s + c^{DL}_2 \left(\alpha_s \ln^2 s\right)^2 + c^{DL}_2 \left(\alpha_s \ln^2 s\right)^3+... \right],
\end{equation}
where $c^{DL}_n$ are numerical factors and $A_{Born}$ denotes amplitude $A$ in the Born approximation. When the resummation
of the series (\ref{dlser}) is done, amplitude $A$ is calculated in DL approximation (DLA).
Total resummation of DL contributions of the graphs involving the single quark loop (see Fig.~1) to process (\ref{gamma})    was
done in Ref.~\cite{bl} for the case of the forward kinematics at $t = 0$.
Calculations in  Ref.~\cite{eit1}  confirmed the results of Ref.~\cite{bl} and generalized them on
the case of  $t \neq 0$, and also accounted for the running coupling effects.

Further progress in studying the process (\ref{gamma})  involves accounting for the
 graphs where the initial $t$-channel photons $\gamma^* (p),\gamma^* (p')$
and the final photons $\gamma^* (q),\gamma^* (q')$ are attached to different quark loops.
These loops are related by an arbitrary intermittence of the $t$-channel quark and gluon ladder rungs complemented
by non-ladder contributions.
An example of the involved graphs is shown in Fig.~2, where non-ladder contributions are not shown for simplicity.
The $u$-channel ladder graph with replacement $p \leftrightarrows p'$ is also not shown.

\begin{figure}[h]
  \includegraphics[width=.2\textwidth]{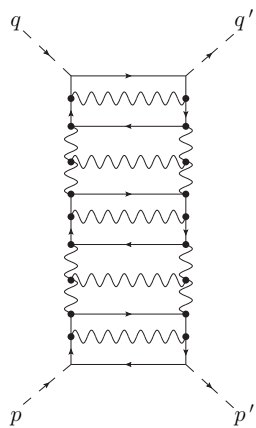}
  \caption{\label{gammaladfig2} Example of the graphs with quark and gluon ladder rungs.}
\end{figure}

Such graphs yield DL contributions, however they also bring the other kind of contributions
called leading logarithms (LL) where the single-logarithmic contributions are multiplied by the power of $s$:

\begin{equation}\label{llser}
A_{LL} = s \left[1 + c^{LL}_1 \alpha_s \ln s + c^{LL}_2 (\alpha_s \ln s)^2 + c^{LL}_3 (\alpha_s \ln s)^3 + ...\right],
\end{equation}
 where $c^{LL}_n$ denote numerical factors. It was shown in Ref.~\cite{ttwu} that DL contributions in Eq.~(\ref{llser})
are absent because the DL contributions from involved Feynman graphs cancel each other.
Values of $c^{LL}_n$ in high orders in $\alpha_s$ are mostly
 determined by
 the gluon ladder-like constructions connecting the upper
 and lowest quark blobs (they are called the impact-factors).
 The LO BFKL equation\cite{lobfkl} sums the leading logarithms to all orders in $\alpha_s$.
 Solution to this equation is found in the form
 of the infinite sum of the high-energy asymptotics, each $\sim s^{\omega_n}$. The asymptotic
 with maximal $\omega_n$ (the leading asymptotics) is
called the BFKL Pomeron. Impact of total resummation of  sub-leading contributions $\sim s \alpha_s (\alpha_s \ln s)^n$ on
the Pomeron intercept was calculated in Ref.~\cite{nlobfkl}.

The fact that  DL contributions multiplied by $s$ cancel
 each other means that $A_{Born}$ in the DL series of Eq.~(\ref{dlser})
 cannot contain contributions $\sim s$ in contrast to the
 series (\ref{llser}). By this reason,  there is
the conventional opinion in the literature that the LL contribution (\ref{llser}) surpasses the DL contribution
(\ref{dlser}) to such an extent that
one can neglect DL contributions (\ref{dlser}) compared to (\ref{llser}).
In the present paper we prove that this opinion is false.
We demonstrate that the high-energy asymptotics of the light-by-light amplitudes in DLA
are of the same order as the BFKL result.
To this end we calculate the amplitude $A_{\gamma \gamma}$ of
the process (\ref{gamma}) in DLA at zero and non-zero values of $t$.
By doing so, we account for both quark and gluon DL contributions
and sum them to all orders in $\alpha_s$. First we keep $\alpha_s$ fixed and then
account for the running coupling effects. Once $A_{\gamma \gamma}$ is known, applying the Saddle-Point
method to $A_{\gamma \gamma}$ allows us to calculate its asymptotics. As this asymptotics bears
the vacuum quantum numbers, it is a new contribution to Pomeron. Throughout the paper we address it as the DL Pomeron.
We  compare it to the BFKL Pomeron. As we know both the  amplitude $A_{\gamma \gamma}$ and its
asymptotics, we can estimate the minimal value
of $s$, where the DL Pomeron,  can  reliably represent  its parent amplitude $A_{\gamma \gamma}$.

In order to perform calculations in DLA, we compose and solve Infra-Red Evolution Equations\footnote{This name was suggested by M.~Krawczyk.}
(IREE) for $\hat{A}_{\gamma \gamma}$.
This approach was suggested by L.N.~Lipatov in Refs.~\cite{l,kl} and generalized on the case of inelastic
processes in Ref.~\cite{elf}. It is based on factorization of partons with minimal transverse momenta first proved by
V.N.~Gribov\cite{g} in QED and then extended to the non-Abelian field theories.
The IREE approach has proved to be an efficient instrument for
total resummation of DL contributions to various processes in QED,QCD and other theories. More info
on this approach can be found in the overviews\cite{egtsum}. In the context
of the light-by-light scattering the IREE approach was used in Ref.~\cite{bl,eit1}.

Essence of this approach is
first to neglect masses of involved quarks so as treat quarks and gluons the same way. Then one should
introduce a cut-off $\mu$ to regulate IR divergences. It is convenient to introduce such cut-off $\mu$ in the
transverse momentum space so that transverse momenta of all virtual partons were greater than $\mu$.
In order to neglect the involved quark masses $m_q$ and guarantee applicability of the Perturbative QCD, $\mu$
should satisfy the obvious restrictions:

\begin{equation}\label{qlambda}
\mu > m_q,~~\mu > \Lambda_{QCD}.
\end{equation}

Otherwise a value of $\mu$ is not fixed, so
one can study evolution with respect to $\mu$ and
construct thereby evolution equations. All IREEs look much simple when written in terms of the Mellin
amplitudes $F_{\gamma \gamma}$. We use the Mellin transform in the standard form:

 \begin{equation}\label{mellin}
A_{\gamma \gamma}(s, Q^2_1, Q^2_2) = \int_{- \imath \infty}^{\imath \infty}
\frac{d \omega}{2 \pi \imath} \left(s/\mu^2\right)^{\omega} F_{\gamma \gamma}(\omega, Q^2_1,Q^2_2)
= \int_{- \imath \infty}^{\imath \infty}
\frac{d \omega}{2 \pi \imath} e^{\omega \rho} F_{\gamma \gamma}(\omega, y_1,y_2),
\end{equation}
where we have introduced the $\mu$-dependent logarithmic variables $\rho, y_1, y_2$:
\begin{equation}\label{y12}
\rho = \ln (s/\mu^2),~~y_1 = \ln (Q^2_1/\mu^2),~~y_2 = \ln (Q^2_2/\mu^2).
 \end{equation}

Calculation of  $A_{\gamma \gamma}$ in DLA leads to different results,
$A^{(M)}_{\gamma \gamma}$  and $A^{(D)}_{\gamma \gamma}$ , depending on relations between
$s, \mu$ and $Q^2_{1,2}$. When the external photons are Moderately-Virtual, i.e. when

\begin{equation}\label{mkin}
s \mu^2 \gg Q^2_{1}Q^2_{2},
\end{equation}
we provide $A_{\gamma \gamma}$ with the superscript M: $A^{(M)}_{\gamma \gamma}$.
The amplitude $A^{(D)}_{\gamma \gamma}$  corresponds to opposite case where

\begin{equation}\label{dkin}
s \mu^2 \ll Q^2_{1}Q^2_{2}.
\end{equation}
We call it the case of  Deeply-Virtual photons. In the present paper we calculate both
$A^{(M)}_{\gamma \gamma}$  and $A^{(D)}_{\gamma \gamma}$. Throughout the paper we will
use the generic notation $A_{\gamma \gamma}$ in the cases insensitive to the difference
between  $A^{(M)}_{\gamma \gamma}$ and $A^{(D)}_{\gamma \gamma}$.

Our paper is organized as follows: in Sect.~II we compose and solve IREE for $A^{(M,D)}_{\gamma \gamma}$,
expressing them in terms of the auxiliary amplitudes $A_{\gamma q}$ and $A_{\gamma g}$ (the superscripts
$q,g$ refer to quarks and gluon respectively). Amplitudes
$A_{\gamma q}, A_{\gamma g}$ are expressed through parton-parton amplitudes in
Sect.~III. In Sect.~IV we obtain explicit expressions for the parton-parton amplitudes and we
arrive at explicit expressions for $A^{(M,D)}_{\gamma \gamma}$. In Sects. II- IV we considered the
particular case of the forward kinematics assuming that $t = 0$. In Sect.~V we
study the photon-photon scattering amplitudes in the forward kinematics with non-zero $t$ and arbitrary relations
between $t$ and $Q^2_{1,2}$. In Sect.VI we calculate the asymptotics
of $A^{(M,D)}_{\gamma \gamma}$ at $s \to \infty$, applying the saddle-point method and
arrive at a new Pomeron. Then we study the applicability region for the use of the asymptotics.
In Sect.~VII we compare the DL Pomeron to the BFKL Pomeron and discuss their
similarities and differences. Finally, Sect.~VIII is for concluding remarks.

\section{Evolution equations for photon-photon scattering amplitudes}

In this Sect. we study the process (\ref{gamma}) in the forward
kinematics (\ref{fkin}) at
$t = 0$. We construct IREEs for the amplitude $A_{\gamma \gamma}(s,Q^2_1, Q^2_2)$. They
represent $A_{\gamma \gamma}(s,Q^2_1, Q^2_2)$
through the amplitudes of photon-parton scattering. For constructing
the IREEs, we use that the $t$-channel two-parton state factorizes the graph in Fig.~2 into
two parts providing the
transverse momentum  (with the plane formed by $p$ and $q$) of the pair
is minimal compared to other transverse momenta. Such partons are called the softest. Technically, constructing IREEs for $A_{\gamma \gamma}$
is pretty similar to the derivation of the IREEs considered
in Ref.~\cite{eit1} with one important exception:  The softest partons in Ref.~\cite{eit1} were quarks while now they can be
both quarks and gluons.
Combining the both these opportunities, we come to the IREE depicted in Fig.~3.

\begin{figure}[h]
  \includegraphics[width=.5\textwidth]{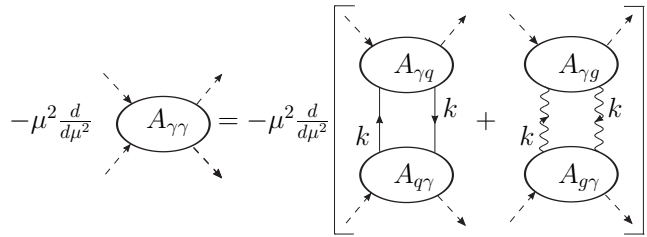}
  \caption{\label{gammaladfig3} IREE for $A_{\gamma \gamma}$. The dashed lines stand for the external photons, the
  straight (wavy) lines denote the quark (gluon) propagators of the softest partons with momenta $k$. }
\end{figure}

Applying the standard Feynman rules to the graphs in Fig.~3 and using the Mellin transform (\ref{mellin}), we write the IREE
in the analytical form in the $\omega$-space:

\begin{equation}\label{eqagammagen}
\omega F_{\gamma \gamma} + \frac{\partial F_{\gamma \gamma}}{\partial y_1} +
\frac{\partial F_{\gamma \gamma} }{\partial y_2} = \frac{1}{8 \pi^2} \left[F_{\gamma q} F_{q\gamma } + F_{\gamma g}F_{g \gamma}\right].
\end{equation}

We have used in Eq.~(\ref{eqagammagen}) the logarithmic variables introduced in Eq.~(\ref{y12}). The l.h.s. of Eq.~(\ref{eqagammagen})
corresponds to differentiation of Eq.~(\ref{mellin}). It
includes the derivatives with respect to $y_{1,2}$ and the result of differentiation of the factor $(s/\mu^2)^{\omega}$.
The r.h.s. of Eq.~(\ref{eqagammagen})   involves the auxiliary photon-parton amplitudes so that amplitude
$A_{\gamma g}$ corresponds in the $t$-channel to the two-photon production of a quark-anti-quark pair

\begin{equation}\label{gammaq}
\gamma^* (p)  \gamma^* (q) \to q (p'_1) \bar{q} (p'_2)
\end{equation}

while amplitude $A_{\gamma g}$ describes production of two gluons

\begin{equation}\label{gammag}
\gamma^* (p)  \gamma^* (q) \to g (p'_1) g (p'_2).
\end{equation}

$A_{\gamma g}$ depends on $\mu$ through variables $\rho, y_1, y_2$,
Technology of solving IREEs for objects with several $\mu$-dependent variables
requires that all such variables should be ordered.
We use the ordering of  Eq.~(\ref{fkin}), complementing it by the restriction
$Q^2_1 \gg Q^2_2$ and arriving thereby at

\begin{equation}\label{rhoy}
\rho > y_1 > y_2
\end{equation}
but we will represent the final expressions for $A_{\gamma \gamma}(s, Q^2_1, Q^2_2)$
in the form, where $Q^2_1$ and $Q^2_2$ are not ordered. Amplitude $A_{\gamma \gamma}$ in the IREE (\ref{eqagammagen})
depends on three variables.  The conventional way to solve such many-parametrical IREEs is
firstly to solve IREEs with the lesser number of variables and then to use such solutions for
specifying the
general solution of  (\ref{eqagammagen}).

\subsection{All photons are nearly on-shell}

We start with calculation of $A_{\gamma \gamma}$
in the simplest kinematics where $Q^2_1 \approx Q^2_2 \lesssim \mu^2$, i.e. when

\begin{equation}\label{q12on}
y_2 = y_1 = 0.
\end{equation}

We denote its $f_{\gamma \gamma}(\omega)$ the Mellin amplitude for the kinematics (\ref{q12on}).
The l.h.s. of the IREE for $f_{\gamma \gamma}(\omega)$ corresponds to the l.h.s. of Eq.~(\ref{eqagammagen})
where the $y_{1,2}$-dependence is absent while
the r.h.s.  coincides with the one of Eq.~(\ref{eqagammagen}).  Therefore we obtain

\begin{equation}\label{fon}
\omega f_{\gamma \gamma}(\omega) = \frac{1}{8 \pi^2} \left[f_{\gamma q}(\omega) f_{q \gamma}(\omega)
+ f_{\gamma g}(\omega) f_{g \gamma}(\omega)\right],
\end{equation}
where amplitudes $f_{\gamma q}(\omega)$ and $f_{q \gamma}(\omega)$ correspond to the processes
(\ref{gammaq}) and the reversed process respectively.
Similarly,  $f_{\gamma g}(\omega)$ and $f_{g \gamma}(\omega)$ correspond to the process
(\ref{gammag}) and the reversed process respectively.

\subsection{One of the photons is on-shell and the other is off-shell }

Let us consider the more complicated case when

\begin{equation}\label{y2mu}
\rho > y_1 > y_2 = 0,
\end{equation}
i.e. $Q^2_1 \gg Q^2_2 \sim \mu^2$.
We denote $\widetilde{A}_{\gamma\gamma} (s, y_1)$ the amplitude corresponding to that case and keep the notation
$\widetilde{F}_{\gamma\gamma} (\omega, y_1)$ for the Mellin amplitude of this process. Again, the r.h.s. of the new IREE coincides
with
the r.h.s. of Eq.~(\ref{eqagammagen}). The l.h.s. corresponds to the l.h.s. of Eq.~(\ref{eqagammagen}) without the second derivative
 because $y_2 = 0$:

\begin{equation}\label{eq2gen}
\frac{\partial \widetilde{F}_{\gamma\gamma}}{\partial y_1}
+ \omega \widetilde{F}_{\gamma\gamma} = \frac{1}{8 \pi^2} F_{\gamma q} (\omega, y_1)
f_{q \gamma} (\omega) + \frac{1}{8 \pi^2} F_{\gamma g} (\omega, y_1)
f_{g \gamma} (\omega).
\end{equation}

Amplitudes $F_{\gamma q} (\omega, y_1), F_{\gamma g} (\omega, y_1)$ and $f_{g \gamma} (\omega), f_{g \gamma} (\omega)$ are supposed to be
calculated independently.
Once they are known, the general solution to Eq.~(\ref{eq2gen}) is

\begin{equation}\label{sol2gen}
 \widetilde{F}_{\gamma\gamma} = e^{-\omega y_1}
\left[C_2 (\omega) + \frac{1}{8 \pi^2}  f_{q \gamma} (\omega) \int_0^{y_1} d y' e^{\omega y'} F_{\gamma q} (\omega, y')
+ \frac{1}{8 \pi^2}  f_{g \gamma} (\omega) \int_0^{y_1} d y' e^{\omega y'} F_{\gamma g} (\omega, y')
\right]
\end{equation}

In order to specify an unknown function $C_2$ in Eq.~(\ref{sol2gen}) we use the matching:

\begin{equation}\label{match2}
\widetilde{F}_{\gamma\gamma} (\omega, y_1)|_{y_1 = 0} = f_{\gamma\gamma} (\omega),
\end{equation}
where $f_{\gamma\gamma} (\omega)$ is defined in Eq.~(\ref{fon}). Therefore,
$\widetilde{F}_{\gamma\gamma} (\omega, y_1)$ in kinematics (\ref{y2mu}) is represented in terms of the photon-quark amplitudes:

\begin{equation}\label{sol2}
 \widetilde{F}_{\gamma\gamma} (\omega, y_1) = \frac{1}{8 \pi^2} e^{-\omega y_1}\left[
 \sum_r \frac{1}{\omega}  f_{\gamma r} (\omega)
 f_{r \gamma} (\omega) + \sum_r
 f_{r \gamma} (\omega) \int_0^{y_1} d y e^{\omega y} F_{\gamma r} (\omega, y)
 \right],
\end{equation}
where $r = q,g$ so that $F_{\gamma r} (\omega, y)$ stands for the photon-quark and photon-gluon amplitudes at $y \neq 0$.

\subsection{Off-shell photons with moderate  virtualities}

We call the moderately virtual kinematics the case when
$Q^2_1 \gg \mu^2$ and $Q^2_2 \gg \mu^2$ but $Q^2_1 Q^2_2 \ll s \mu^2$. In the logarithmic
variables it means that

\begin{equation}\label{modoffshell}
\rho > y_2 + y_1.
\end{equation}

The IREE for the amplitude $F_{\gamma\gamma}^{(M)} (\omega, y_1, y_2)$  in the kinematic region (\ref{modoffshell}) is

\begin{equation}\label{eq12gen}
\frac{\partial F_{\gamma\gamma}^{(M)}}{\partial y_2} + \frac{\partial F_{\gamma\gamma}^{(M)}}{\partial y_1}
+ \omega F_{\gamma\gamma}^{(M)} = \frac{1}{8 \pi^2}
\sum_r
F_{\gamma r} (\omega, y_1) F_{r \gamma} (\omega, y_2).
\end{equation}

In order to use the symmetry with respect to $y_1, y_2$ of the differential operator
in (\ref{eq12gen}) and simplify the IREE, we have introduced new variables $\xi, \eta$:

\begin{equation}\label{xieta}
\xi = y_1 + y_2,~~\eta = y_1 - y_2.
\end{equation}

Eq.~(\ref{eq12gen}) in terms of $\xi, \eta$ takes a simpler form:

\begin{equation}\label{eq12xieta}
2 \frac{\partial F_{\gamma\gamma}^{(M)}}{\partial \xi}
+ \omega F_{\gamma\gamma}^{(M)} = \frac{1}{8 \pi^2} \sum_r
F_{\gamma r} \left(\omega, y_1 \right)
F_{r \gamma} \left(\omega, y_2\right).
\end{equation}

A general solution to Eq.~(\ref{eq12xieta})
is

\begin{equation}\label{sol12gen}
F_{\gamma\gamma}^{(M)} = e^{- \omega \xi/2} \left[C(\omega, \eta) + \frac{1}{16 \pi^2}
\sum_r
\int_0^{\xi} d \xi'
e^{\omega \xi'/2} F_{\gamma r} (\omega, y'_1) F_{r \gamma} (\omega, y'_2) \right],
\end{equation}
with $C(\omega, \eta)$ being an arbitrary function and the variables $y'_1,y'_2$ are defined as
follows:

\begin{equation}\label{y12prime}
y'_1 = (\xi' + \eta)/2,~~y'_2 = (\xi' -\eta)/2.
\end{equation}

In order to specify $C(\omega, \eta)$, we use the matching of $F_{\gamma\gamma}^{(M)} (\omega, y_1,y_2)$
with an amplitude $\widetilde{F}_{\gamma\gamma} (\omega, y_1)$
of the same process but in the simpler kinematic regime (\ref{y2mu}) considered above:

\begin{equation}\label{match1}
F_{\gamma\gamma}^{(M)} (\omega, y_1,y_2)|_{y_2 = 0} = \widetilde{F}_{\gamma\gamma} (\omega, y_1),
\end{equation}

Combining Eqs.~(\ref{match1}), (\ref{sol12gen}) and (\ref{sol2}), we arrive at the following expression for
$F_{\gamma\gamma}$:

\begin{eqnarray}\label{fm12}
F_{\gamma\gamma}^{(M)} &=& e^{- \omega \xi/2} \left[e^{\omega \eta/2}
\widetilde{F}_{\gamma\gamma} (\omega, \eta)
+ \frac{1}{16 \pi^2}
\sum_r
\int_{\eta}^{\xi} d \xi' e^{\omega \xi'/2}
F_{\gamma r} (\omega, y'_1) F_{r \gamma} (\omega, y'_2)\right].
\end{eqnarray}

Eq.~(\ref{fm12}) is obtained under assumption that $y_1 > y_2$ (see Eq.~(\ref{rhoy})).
Replacing  in Eq.~(\ref{fm12})
$\eta$ by $|\eta|$, allows us to describe the cases $y_1 > y_2$ and $y_1 < y_2$ at the same time:

\begin{eqnarray}\label{fm}
F_{\gamma\gamma}^{(M)} &=& e^{- \omega \xi/2} \left[e^{\omega |\eta|/2}
\widetilde{F}_{\gamma\gamma} (\omega, |\eta|)
+ \frac{1}{16 \pi^2}
\sum_r
\int_{|\eta|}^{\xi} d \xi' e^{\omega \xi'/2}
F_{\gamma r} (\omega, y'_1) F_{r \gamma} (\omega, y'_2)\right].
\end{eqnarray}

Substituting Eq.~(\ref{fm}) in (\ref{mellin}), we arrive at the expression for
the amplitude $A_{\gamma\gamma}^{(M)}$ at moderate virtualities $Q^2_{1,2}$:

\begin{eqnarray}\label{am}
A_{\gamma\gamma}^{(M)} &=&
\int_{- \imath \infty}^{\imath \infty}
\frac{d \omega}{2 \pi \imath}e^{\omega (\rho- \xi/2)}
\left[e^{\omega |\eta|/2}
\widetilde{F}_{\gamma\gamma} (\omega, |\eta|)
+ \frac{1}{16 \pi^2}
\sum_r
\int_{|\eta|}^{\xi} d \xi' e^{\omega \xi'/2}
F_{\gamma r} (\omega, y'_2) F_{r \gamma} (\omega, y'_1)\right]
\\ \nonumber
&=& \int_{- \imath \infty}^{\imath \infty}
\frac{d \omega}{2 \pi \imath}
\left(\frac{s}{\sqrt{Q^2_1 Q^2_2}}\right)^{\omega}
\left[e^{\omega |\eta|/2}
\widetilde{F}_{\gamma\gamma} (\omega, |\eta|)
+ \frac{1}{16 \pi^2}
\sum_r
\int_{|\eta|}^{\xi} d \xi' e^{\omega \xi'/2}
F_{\gamma r} (\omega, y'_2) F_{r \gamma} (\omega, y'_1)\right].
\end{eqnarray}

\subsection{Deeply-virtual photons}

When $Q^2_1 >> \mu^2$ and $Q^2_2 >> \mu^2$, and their product is also great, $Q^2_1 Q^2_2 >> s \mu^2$,
the inequality in Eq.~(\ref{modoffshell})
is replaced by the opposite one

\begin{equation}\label{deepy}
\rho < y_2 + y_1.
\end{equation}

We address such photons as deeply-virtual ones. The principal difference between this case and the case of moderately-virtual
photons is that the scattering amplitude $A_{\gamma\gamma}^{(D)} (\rho,y_1,y_2)$ in the kinematics (\ref{deepy})
does not depend on $\mu$, so the IREE for it is very simple:

\begin{equation}\label{eqad}
\frac{\partial A_{\gamma\gamma}^{(D)}}{\partial \rho}
+ \frac{\partial A_{\gamma\gamma}^{(D)}}{\partial y_1 } +
\frac{\partial A_{\gamma\gamma}^{(D)}}{\partial y_2} = 0.
\end{equation}

A general solution to Eq.~(\ref{eqad}) can be written in different ways. The most convenient
way for us is

\begin{equation}\label{adeepgen}
A_{\gamma\gamma}^{(D)} = A_{\gamma\gamma}^{(D)} (\rho - y_1, \rho - y_2),
\end{equation}
In order to specify $A_{\gamma\gamma}^{(D)}$ we
use the matching with the amplitude $A_{\gamma\gamma}^{(M)}$ of the same process but in the region
(\ref{modoffshell}). It means that

\begin{equation}\label{match12}
A_{\gamma\gamma}^{(D)} (\rho-y_1, \rho - y_2)|_{\rho = y_1 + y_2}
= A_{\gamma\gamma}^{(M)}(\rho,y_1,y_2)|_{\rho = y_1 + y_2}
\equiv \widetilde{A}_{\gamma\gamma}^{(M)}(y_1,y_2).
\end{equation}

Replacing $y_1 \to \rho - y_2$ and  $y_2 \to \rho - y_1$  in $\widetilde{A}_{\gamma\gamma}^{(M)}(y_1,y_2)$
immediately allows us to express $A_{\gamma\gamma}^{(D)}$
through $\widetilde{A}_{\gamma\gamma}^{(M)}$ in the whole the region $\rho \leq y_1 + y_2$:

\begin{equation}\label{adam}
A_{\gamma\gamma}^{(D)} (\rho,y_1,y_2) = \widetilde{A}_{\gamma\gamma}^{(M)}(\rho - y_2, \rho - y_1).
\end{equation}

Using Eqs.~(\ref{am},\ref{adam}) allows us to write the explicit expression for $A_{\gamma\gamma}^{(D)}$:

\begin{equation}\label{ad}
A_{\gamma\gamma}^{(D)} =
\int_{- \imath \infty}^{\imath \infty}
\frac{d \omega}{2 \pi \imath}e^{\omega (\rho- \xi/2)}
\left[e^{\omega |\eta|/2}
\widetilde{F}_{\gamma\gamma} (\omega, |\eta|)
+ \frac{1}{16 \pi^2}
\sum_r
\int_{|\eta|}^{2 \rho -\xi} d \xi' e^{\omega \xi'/2}
F_{\gamma r} (\omega, y'_2) F_{r \gamma} (\omega, y'_1)\right]
\end{equation}

\section{Amplitudes of photon-parton scattering}

\subsection{Photon-parton amplitude with on-shell photon}

We consider the case when $y = 0$ and denote $f_{\gamma q}(\omega)$ and $f_{\gamma g}(\omega)$
the Mellin amplitudes of processes (\ref{gammaq}) and (\ref{gammag}) respectively.
In the $\omega$-space the system of IREE is

\begin{eqnarray}\label{eqfy0}
 \omega f_{\gamma q}(\omega) &=&
a_{\gamma q} +
f_{\gamma q} (\omega) h_{qq} (\omega) + f_{\gamma g} (\omega) h_{gq} (\omega),
\\ \nonumber
\omega f_{\gamma g}(\omega) &=&
f_{\gamma q} (\omega) h_{qg} (\omega) + f_{\gamma g} (\omega) h_{gg} (\omega),
\end{eqnarray}
where $a_{\gamma q} = e^2$, so that $a_{\gamma q}/\omega$ is the Born value of amplitude $f_{\gamma q}(\omega)$.
There is no a similar term in the equation for $f_{\gamma g}(\omega)$.
We have used the following convenient notations in the rhs of Eq.~(\ref{eqfy0}):
\begin{equation}\label{fh}
h_{rr'} = \frac{1}{8 \pi^2} f_{rr'},
\end{equation}
with $r,r' = q,g$ and $f_{rr'}$ being the parton-parton amplitudes. The solution to Eq.~(\ref{eqfy0}) is

\begin{eqnarray}\label{fy0}
 f_{\gamma q} (\omega) &=& \frac{a_{\gamma q} (\omega - h_{gg})}{K(\omega)},
 \\ \nonumber
f_{\gamma q} (\omega) &=&  \frac{a_{\gamma q} h_{qg}}{K(\omega)},
\end{eqnarray}
with $K(\omega)$ being the determinant of the system (\ref{eqfy0}):

\begin{equation}\label{k}
K = (\omega - h_{qq})(\omega - h_{gg}) - h_{gg}h_{qg}.
\end{equation}

Amplitudes $f_{r \gamma}$ are given by expressions similar to Eq.~(\ref{fy0}). They can be
obtained from (\ref{fy0}) with replacement
$a_{\gamma q}$ by $a_{q\gamma } = - e^2$, which means that

\begin{equation}\label{frgamma}
f_{q \gamma } = -f_{\gamma q},~~f_{g \gamma }= - f_{g \gamma}.
\end{equation}

\subsection{Off-shell photons}

When the photons are off-shell, the system is more involved:

\begin{eqnarray}\label{eqfy}
\left[\partial/\partial y + \omega \right] F_{\gamma q}(\omega,y) &=&
F_{\gamma q} (\omega, y) h_{qq} (\omega) + F_{\gamma g} (\omega, y) h_{gq} (\omega),
\\ \nonumber
\left[\partial/\partial y + \omega \right] F_{\gamma g}(\omega, y) &=&
F_{\gamma q} (\omega, y) h_{qg} (\omega) + F_{\gamma g} (\omega, y) h_{gg} (\omega)
\end{eqnarray}
and a general solution to it is

\begin{eqnarray}\label{fy}
F_{\gamma q}(\omega,y) &=& e^{- \omega y} \left[C_1 e^{\lambda_1 y} + C_2 e^{\lambda_2 y}\right],
\\ \nonumber
F_{\gamma g}(\omega, y) &=& e^{- \omega y} \left[C_1 \frac{\lambda_1 - h_{qq}}{ h_{qg}}e^{\lambda_1 y} +
C_2  \frac{\lambda_2 - h_{qq}}{ h_{qg}} e^{\lambda_2 y}\right],
\end{eqnarray}

where
%

\begin{equation}\label{omegapm}
\lambda_{1,2} = \frac{1}{2} \left[h_{gg} - h_{qq} \pm \sqrt{R}\right]
\end{equation}
and

\begin{equation}\label{r}
R = (h_{gg} + h_{qq})^2 - 4(h_{qq}h_{gg} - h_{qg}h_{gq}) = (h_{gg} - h_{qq})^2  + 4 h_{qg}h_{gq} .
\end{equation}

We specify the factors $C_{1,2} (\omega)$  by the matching with the solution (\ref{fy0}) of Eq.~(\ref{eqfy0}):

\begin{equation}\label{matchfy0}
F_{\gamma q}(\omega,y)|_{y = 0} = f_{\gamma q}(\omega),~~F_{\gamma g}(\omega,y)|_{y = 0} = f_{\gamma g}(\omega),
\end{equation}
with $f_{\gamma q}$ and $f_{\gamma g}$ being calculated in Eq.~(\ref{fy0}).
This matching leads us to the following expressions:

\begin{eqnarray}\label{cpm}
C_1 &=&  \frac{f_{\gamma g} h_{qg} - f_{\gamma q}(\lambda_2 - h_{qq})}{\sqrt{R}},
\\ \nonumber
C_2 &=& \frac{- f_{\gamma g} h_{qg} - f_{\gamma q}(\lambda_1 - h_{qq})}{\sqrt{R}} .
\end{eqnarray}

Now the r.h.s. of Eqs.~(\ref{fy}) are expressed in terms of the parton-parton amplitudes $h_{rr'}$.
Expressions for amplitudes $F_{r \gamma}$ can be obtained from Eq.~(\ref{fy}) with replacement
$f_{\gamma r}$ by $f_{r \gamma}$ defined in  Eq.~(\ref{frgamma}).

\section{Amplitudes of parton-parton scattering}

Amplitudes $h_{rr'}$ satisfy the system of the algebraic IREEs:

\begin{eqnarray}\label{eqh}
\omega h_{qq} &=& b_{qq} + h_{qq}h_{qq} + h_{qg}h_{gq},~~\omega h_{qg} = b_{qg} + h_{qq}h_{qg}+ h_{qg}h_{gg},
\\ \nonumber
\omega h_{gq} &=& b_{gq} + h_{gq}h_{qq}+ h_{gg}h_{gq},~~\omega h_{gg} = b_{gg} + h_{gq}h_{qg}+ h_{gg}h_{gg},
\end{eqnarray}
where the inhomogeneous terms $b_{ik}$  consist of two contributions:

\begin{equation}\label{abv}
b_{ik} = a_{ik} + V_{ik},
\end{equation}
with $a_{ik}$ being the Born amplitudes divided by the factor $(8 \pi^2)$. They were calculated in Refs.~\cite{ber,egtg1} :

\begin{equation}\label{aik}
a_{qq} = \frac{A(\omega)C_F}{2\pi},~a_{qg} = \frac{A'(\omega)C_F}{\pi},~a_{gq} = -\frac{A'(\omega)n_f}{2 \pi}.
~a_{gg} = \frac{2N A(\omega)}{\pi},
\end{equation}
where $A$ and $A'$ stand for the running QCD couplings:

\begin{eqnarray}\label{a}
A = \frac{1}{b} \left[\frac{\eta}{\eta^2 + \pi^2} - \int_0^{\infty} \frac{d z e^{- \omega z}}{(z + \eta)^2 + \pi^2}\right],
A' = \frac{1}{b} \left[\frac{1}{\eta} - \int_0^{\infty} \frac{d z e^{- \omega z}}{(z + \eta)^2}\right],
\end{eqnarray}
with $\eta = \ln \left(\mu^2/\Lambda^2_{QCD}\right)$ and $b$ being the first coefficient of the Gell-Mann- Low function. When the running effects for the QCD coupling
are neglected,
$A(\omega)$ and $A'(\omega)$ are replaced by $\alpha_s$. The terms $V_{rr'}$ are represented in a similar albeit more involved way (see Ref.~\cite{egtg1} for detail):

\begin{equation}
\label{vik} V_{rr'} = \frac{m_{rr'}}{\pi^2} D(\omega)~,
\end{equation}
with
\begin{equation}
\label{mik} m_{qq} = \frac{C_F}{2 N}~,\quad m_{gg} = - 2N^2~,\quad
m_{gq} = n_f \frac{N}{2}~,\quad m_{qg} = - N C_F~,
\end{equation}
and
\begin{equation}
\label{d} D(\omega) = \frac{1}{2 b^2} \int_{0}^{\infty} d z
e^{- \omega z} \ln \big( (z + \eta)/\eta \big) \Big[
\frac{z + \eta}{(z + \eta)^2 + \pi^2} - \frac{1}{z +
\eta}\Big]~.
\end{equation}

Let us note that $D = 0$ when the running coupling effects are neglected. It corresponds the total compensation of DL
contributions of non-ladder Feynman graphs to scattering amplitudes. When $\alpha_s$ is running, such compensation is only partial.
 Solution to Eq.~(\ref{eqh}) is

\begin{eqnarray}\label{h}
h_{qq} &=& \frac{1}{2} \left[\omega - Z + \frac{(a_{qq}- a_{gg})}{Z}\right],~~h_{qg} = \frac{a_{gq}}{Z},
\\ \nonumber
h_{qg} &=& \frac{a_{gq}}{Z},~~h_{gg} = \frac{1}{2} \left[\omega - Z - \frac{(a_{qq}- a_{gg})}{Z}\right],
\end{eqnarray}
with

\begin{equation}\label{z}
Z = \frac{1}{\sqrt{2}} \left[\omega^2 - 2(a_{qq} + a_{gg}) + W \right]^{1/2},
\end{equation}
with

\begin{equation}\label{w}
W = \sqrt{(\omega^2 - 2a_{qq} -2 a_{gg})^2 - 4(a_{qq}- a_{gg})^2
- 16 a_{qg}a_{gq}}~.
\end{equation}

Eq.~(\ref{h}), with $Z$ given by Eq.~(\ref{z}), presents the explicit expressions for the amplitudes $h_{rr'}$.
Substituting them in Eqs.~(\ref{omegapm},\ref{cpm}) and then in Eqs.~(\ref{fy0},\ref{fy}), we obtain explicit expressions for
the photon-parton amplitudes.   Finally, substituting $F_{\gamma r}$ and $F_{r \gamma}$
in Eqs.~(\ref{sol2},\ref{fm}), we arrive at explicit expressions for the photon-photon amplitudes
$A_{\gamma\gamma}^{(M)},A_{\gamma\gamma}^{(D)}$. As these expressions are cumbersome, we do not write them here.

\section{Photon-photon amplitudes at $t \neq 0$}

In this Sect. we generalize the expressions for $A_{\gamma \gamma}$ on the case of the forward kinematics (\ref{fkin}) where $t \neq 0$.
Technically, this generalization is pretty similar
to what we have done in Ref.~\cite{eit1}. To show it, let us consider  Fig.~3. We start by considering it at $t = 0$.
The r.h.s. of the IREE in Fig.~3 involves two convolutions, where the vertical lines stand for propagators of the $t$-channel partons.
At $t = 0$ each of the propagators is $\sim 1/k^2$ and dealing with polarizations brings the factor $k^2_{\perp}$.
After integrating over the longitudinal components of $k$ has been performed, the propagators yield the factor

\begin{equation}\label{teq0}
\sim \frac{k^2_{\perp}}{k^2_{\perp} k^2_{\perp}},
\end{equation}

with $k_{\perp}$ being the component of $k$ transverse to the plane formed by the external momenta.  After cancelling $k^2_{\perp}$ in
Eq.~(\ref{teq0}),
we are left with the factor $1/k^2_{\perp}$ which yields a logarithmic contribution when integrated over $k_{\perp}$. We remind here that, by definition
of $\mu$,
the integration region is

\begin{equation}\label{regmu}
k^2_{\perp} \gg \mu^2 .
\end{equation}

When $t \neq 0$, one of the propagators on each graph in Fig.~3 remains to be $\sim 1/k^2$ while the other becomes $\sim 1/(q-k)^2$, where momentum
$q$ is related to $t$: $-t = q^2$. After integrating over
the longitudinal component of $k$ has been done, the propagators are

\begin{equation}\label{tneq0}
 \sim \frac{k^2_{\perp}}{k^2_{\perp} (q - k)^2_{\perp}} \approx \frac{k^2_{\perp}}{k^2_{\perp} (|t| + k^2_{\perp})}.
\end{equation}

After cancellation of the factors $k^2_{\perp}$, the remaining factor $1/(|t| + k^2_{\perp})$ in Eq.~(\ref{tneq0}) can yield a logarithmic
contribution in the integration region

\begin{equation}\label{regt}
k^2_{\perp} \gg |t|.
\end{equation}

Eq.~(\ref{regt}) means that $|t|$ acts as a new IR cut-off. This is true for the both quark and gluon convolutions in Fig.~3 and
remains unchanged when the impact of the blobs in the convolutions
is accounted for because the blobs bring logarithmic contributions only.

When $t$ is so small that $|t| \leq \mu^2$,
the restriction (\ref{regt}) does not change the region (\ref{regmu}), so there is
no $t$-dependence of the scattering amplitudes and the expressions for $A^{(M,D)}_{\gamma\gamma}$ obtained for $t = 0$ can also
be used at such small $t$.

The case of large $t$, where $-t > \mu^2$,  is less trivial because there is an interplay
between parameters $Q^2_{1,2}$ and $t$.  New photon-photon and photon-parton amplitudes in this case explicitly depend on $t$.
We denote them $M_{\gamma\gamma}$ and $M_{\gamma r}$ ($r = q,g$) respectively.
 In this case $|t|$ really plays the role of a new IR cut-off, so the amplitudes $M_{\gamma\gamma}, M_{\gamma r}$ do not depend on $\mu$ at all.
All expressions obtained in the previous Sects. remains true after replacement  $\mu^2$ by $|t|$. It leads to the following
redefinitions of the variables $\rho, y_1, y_2$:

\begin{equation}\label{redef}
\rho \to \bar{\rho} = \ln \left(s/|t|\right), y_1 \to \bar{y}_1 = \ln \left(Q^2_1/|t|\right),
y_2 \to \bar{y}_2 = \ln \left(Q^2_2/|t|\right),
\bar{\xi} = \bar{y}_1 + \bar{y}_2, \bar{\eta} = \bar{y}_1 - \bar{y}_2.
\end{equation}

After this redefinition, the amplitudes $A_{\gamma\gamma} (s, y_1,y_2)$ can easily be used to obtain explicit
expressions for $M_{\gamma\gamma} (s, t, y_1, y_2)$ . Let us do it, considering step by step all possible relations between
$t$ and $Q^2_{1,2}$.

\subsection{Case of $Q^2_1 \gg|t| \gg Q^2_2$ or vice versa}

In this case the photon-photon amplitude $\widetilde{M}_{\gamma\gamma}$ does not depend on $Q^2_2$ under the DL accuracy, and is given
by the following expression:

\begin{equation}\label{mtilde}
\widetilde{M}_{\gamma\gamma} (s,t,Q^2_1) = \widetilde{A}_{\gamma\gamma} (\bar{\rho}, \bar{y}_1),
\end{equation}
where $\widetilde{A}_{\gamma\gamma} (\rho, y_1)$ is the amplitude is given by Eqs.~(\ref{mellin},\ref{sol2}).

\subsection{Case of moderately large $t$: $|t| \leq Q^2_{1,2}$}

This case embraces two possibilities. For the case of Moderate-Virtual photons where

\begin{equation}\label{mvt}
s |t| \gg Q^2_1 Q^2_2,
\end{equation}
the photon-photon amplitude $M^{(M)}_{\gamma\gamma} (s, t, y_1, y_2)$ is given by the expression
of Eq.~(\ref{am}) combined with (\ref{redef}):

\begin{equation}\label{mm}
M^{(M)}_{\gamma\gamma} (s, t, y_1, y_2) = A^{(M)}_{\gamma\gamma} (\bar{\rho}, \bar{\xi}, \bar{\eta}).
\end{equation}

In the opposite case of Deeply-Virtual photons obeying

\begin{equation}\label{mdt}
s |t| \ll Q^2_1 Q^2_2,
\end{equation}

the photon-photon amplitude $M^{(D)}_{\gamma\gamma} (s, t, y_1, y_2)$ is expressed though amplitude $A^{(D)}_{\gamma\gamma}$
of Eq.~(\ref{ad}):

\begin{equation}\label{md}
M^{(D)}_{\gamma\gamma} (s, t, y_1, y_2) = A^{(D)}_{\gamma\gamma} (\bar{\rho}, \bar{\xi}, \bar{\eta}).
\end{equation}

\subsection{Case of very large $t:|t| \gg Q^2_{1,2}$}

In this case the photon-photon amplitude is insensitive to $Q^2_{1,2}$ under the DL accuracy, so amplitude
$M^{(0)}_{\gamma\gamma}$ is

\begin{equation}\label{mon}
M^{(D)}_{\gamma\gamma} (s,t) = A^{(0)}_{\gamma\gamma} (\bar{\rho}),
\end{equation}
where amplitude $A^{(0)}_{\gamma\gamma} (\rho)$ is given by  Eqs.~(\ref{mellin},\ref{fon}).

Replacing $s$ by $u$ in the expressions considered in the cases above, we obtain amplitudes $M'_{\gamma\gamma}$. Adding
them to $M_{\gamma\gamma}$, we eventually arrive at amplitudes $\hat{M}_{\gamma\gamma}$ describing the photon-photon scattering
in the forward kinematics at $t \neq 0$.

\section{Asymptotics of the light-by-light amplitudes}

When $s \to \infty$ and $Q^2_{1,2}$ are fixed, the amplitudes $A_{\gamma\gamma}^{(M)},A_{\gamma\gamma}^{(D)}$
can be represented by their asymptotic expressions which, strictly speaking, should be obtained by using the saddle-point
method.
There is no difference between treatment of $A_{\gamma\gamma}^{(M)}$ and
$A_{\gamma\gamma}^{(D)}$
to obtain their asymptotics, so we skip the superscripts $M,D$ in what follows.

\subsection{Application of the saddle-point method to amplitudes $A\varpi$}

To begin with, we represent
$A_{\gamma\gamma}$  in the form

\begin{equation}\label{ampsi}
 A_{\gamma\gamma} =
\int_{- \imath \infty}^{\imath \infty}
\frac{d \omega}{2 \pi \imath}e^{\omega \rho + \Psi},
\end{equation}
\begin{equation}\label{psi}
\Psi \approx - \omega \xi/2 + \ln F.
\end{equation}
Now one can expand the exponent in Eq.~(\ref{ampsi}) in the power series at $\omega = \omega_0$, retaining the first three terms:
\begin{equation}\label{psiser}
\omega \rho + \Psi (\omega) \approx \omega_0 \rho + \Psi (\omega_0) + (\rho + \Psi'(\omega_0)) (\omega - \omega_0) + (1/2)
\Psi^{''}(\omega_0) (\omega - \omega_0)^2
\end{equation}
and solve the stationary point  equation

\begin{equation}\label{eqomega0gen}
\rho + \Psi' = 0.
\end{equation}

The rightmost root of this equation, $\omega_0$ is the leading
singularity.
After that $A_{\gamma\gamma}^{(M)}$ can be represented as follows:
\begin{equation}\label{asamgen}
A_{\gamma\gamma} \sim A_{\gamma\gamma}^{as} = \frac{e^{\Psi(\omega_0)}}{\sqrt{2 \pi {\Psi'}' (\omega_0,\rho)}}\left(\frac{s}{\mu^2}\right)^{\omega_0}
= \frac{F(\omega_9)}{\sqrt{2 \pi {\Psi'}' (\omega_0,\rho)}}
\left(\frac{s}{\sqrt{Q^2_1 Q^2_2}}\right)^{\omega_0}.
\end{equation}

Let us notice in advance that  it follows from Eqs.~(\ref{eqomega0},\ref{omegapm},\ref{h})
that $\lambda_1 (\omega_0) > \lambda_2 (\omega_0)$
and therefore $\lambda_2(\omega_0)$ can be dropped in asymptotic expressions.
Eq.~(\ref{eqomega0gen}) can be written as follows:

\begin{equation}\label{eqomega}
\rho - \xi/2  + F'/F= 0.
\end{equation}

In Eq.~(\ref{eqomega}) $\rho \to \infty$, so it must be equated by another large negative contribution.
The only option is that such term corresponds to a singularity of $F'$.
Writing

\begin{equation}\label{d1w}
\frac{F'}{F} = \frac{1}{F} \frac{\partial F}{\partial W} \frac{d W}{d \omega} =
\frac{1}{F} \frac{\partial F}{\partial W} \frac{\lambda}{W}
= \chi \frac{\lambda}{W},
\end{equation}
where we have denoted

\begin{equation}\label{chi}
\chi = \frac{1}{F} \frac{\partial F}{\partial W},~~ \lambda = 2 \omega (\omega^2 - 2b_{qq} -2 b_{gg}),
\end{equation}

and using the explicit expression for $F$, we see that the rightmost
singularity is $W = 0$, so that the leading singularity is the largest root of the equation

\begin{equation}\label{eqomega0}
 (\omega^2 - 2b_{qq} -2 b_{gg})^2 - 4(b_{qq}- b_{gg})^2
- 16 b_{qg}b_{gq} = 0.
\end{equation}

In vicinity of $\omega_0$ Eq.~(\ref{eqomega}) is

\begin{equation}\label{eqomegarho}
\rho + \chi (\omega_0)\frac{\lambda (\omega_0)}{W_0} = 0,
\end{equation}
so

\begin{equation}\label{w0}
W_0 = - \chi (\omega_0)\frac{\lambda (\omega_0)}{\rho}.
\end{equation}

The main contribution to $\Psi^{''}$ comes from differentiation of the factor $1/W$ in Eq.~(\ref{d1w}), so

\begin{equation}\label{psi2}
\Psi^{''} (\omega_0) \approx - \chi(\omega_0) \frac{\lambda^2(\omega_0)}{W^3_0} =
\frac{\rho^3}{\chi^2(\omega_0)\lambda(\omega_0)} \equiv \sigma \rho^3.
\end{equation}

Substituting it in Eq.~(\ref{asamgen}), we obtain the explicit expression for the asymptotics $A_{\gamma\gamma}^{as}$ of
amplitude $A_{\gamma\gamma}$:
When $s \to \infty$ while $Q^2_{1,2}$ are fixed, $A_{\gamma\gamma} \to A_{\gamma\gamma}^{as}$, where

\begin{equation}\label{asam}
A_{\gamma\gamma}^{as} = \frac{F(\omega_0)}
{\sqrt{2 \pi \sigma \rho^3}}
\left(\frac{s}{\sqrt{Q^2_1 Q^2_2}}\right)^{\omega_0}.
\end{equation}

Eq.~(\ref{asam}) manifests that the asymptotics of $A_{\gamma\gamma}^{(M,D)}$ are
of the Regge form, with the exponential being a Reggeon. As its quantum numbers in the $t$-channel are vacuum ones,
it is a new contribution to Pomeron. we will address it as  DL Pomeron. It universally (save the factors $F(\omega_0)$ and $\sigma$) contributes
to high-energy asymptotics of all QCD processes where vacuum quantum numbers are allowed.
Eq.~(\ref{eqomega0}) was first obtained in the context of the small-$x$ asymptotics of the
DIS structure function $F_1$ in
Ref.~\cite{etf1}. Let us discuss Eq.~(\ref{eqomega0}) in more detail.
First of all, we notice that Eq.~(\ref{eqomega0}) can be solved analytically only under the approximation of fixed $\alpha_s$.
In this case the rightmost solution of Eq.~(\ref{eqomega0}) is

\begin{equation}\label{eqomega0fix}
\omega_0^{fix} =
\left(\frac{\alpha_s}{\pi}\right)^{1/2} \left[4 N + C_F + \sqrt{(4N - C_F)^2 - 8 n_f C_F}\right]^{1/2}.
\end{equation}

In order to fix $\alpha_s$ in Eq.~(\ref{eqomega0fix}), we use the value $\alpha_s = 0.24$ obtained in Ref.~(\cite{egtfix}).
When the quark contribution to $\omega_0$ is neglected, we obtain

\begin{equation}\label{omega0fixg}
\omega_0^{(fix~g)} = 1.35.
\end{equation}

Accounting for quark contribution at $n_f = 4$ diminishes value of $\omega_0$. In this case we obtain

\begin{equation}\label{omega0fix}
\omega_0^{fix} = 1.29.
\end{equation}

 However, the approximation of fixed $\alpha_s$  inevitably involves a procedure of setting
 $\alpha_s$ in the final expressions, which is far from being rigorous (see e.g. Refs.~\cite{egtfix,kim}). By this reason, the
 fixed $\alpha_s$ approximation can be regarded as too
rough to be conclusive.
In the more realistic case where $\alpha_s$ runs in every vertex of each involved Feynman graph,
Eq.~(\ref{eqomega0})  can be solved only numerically, which was done in Ref.~\cite{etf1}.
When the quark contribution are neglected and $\alpha_s$ runs we obtain

\begin{equation}\label{omega0g}
\omega_0^{(g)} = 1.25.
\end{equation}

At last, accounting for both gluon and quark contribution together with accounting for the running coupling effects, we arrive at

\begin{equation}\label{delta}
\omega_0 = 1.066.
\end{equation}

Eqs.~(\ref{omega0fixg}-\ref{delta}) demonstrate that DL Pomeron is always supercritical though the intercept is decreasing when
accuracy of the calculations increases.
The Regge asymptotics always look much simpler than the parent amplitudes (e.g. cf. $A^{as}_{\gamma\gamma}$ to
 $A_{\gamma\gamma}$) and by this reason they are often used even at the energies pretty far from asymptotic.
 On the other hand, it is not always possible to outline applicability regions where the asymptotics can be used.
 Fortunately, we have such possibility.
 In order to fix the applicability region for the asymptotics (\ref{asam}), we
 construct $R_{as}$ defined as the ratio of the asymptotics versus the parent amplitude:

 \begin{equation}\label{ras}
 R_{as} = A^{as}_{\gamma\gamma}/A_{\gamma\gamma}
 \end{equation}
 and study
 dependence of $ R_{as}$ on $s$ at fixed $Q^2_{1,2}$.
 Obviously, the asymptotics $A^{as}_{\gamma\gamma}$ reliably represents the parent amplitude $A_{\gamma\gamma}$
 when $ R_{as}$ is not far from $1$. In Ref.~\cite{etf1} we showed that $ R_{as}$ grows with $s$, so we suggest that
 the asymptotics can be used at energies $s > s_{min}$ where $s_{min}$ is the energy where $R_{as} \approx 0.9$.
 Numerical estimates show that

 \begin{equation}\label{smin}
 s_{min} = 10^6 \sqrt{Q^2_1 Q^2_2}.
 \end{equation}

At $s \leq s_{min}$ the parent amplitudes (\ref{am},\ref{ad}) should be used instead of their asymptotics (\ref{asam}).

\section{Comparison to the BFKL approach}

The graphs contributing to $A_{\gamma \gamma}$
are represented in  Fig.~4  in the same form as used in the BFKL: the upper and lowest
blobs,
$\Phi_A (p, p',k_A, k'_A)$ and $\Phi_B (k_B,k'_B,q,q')$, are connected by
the middle blob corresponding to the gluon-gluon scattering amplitude $A_{gg} (k_A,k'_A, k_B,k'_B)$.
The blobs $\Phi_{A,B}$ are called impact-factors.
By doing so, we  bring $A_{\gamma\gamma}$ to the form of BFKL-like convolution:

\begin{equation}\label{bfklconv}
A_{\gamma\gamma} = \Phi_A (p, p',k_A, k'_A)\otimes A_{gg} (k_A,k'_A, k_B,k'_B) \otimes \Phi_B (k_B,k'_B,q,q').
\end{equation}

The representation (\ref{bfklconv})
dates back to the phenomenological Regge theory where impact-factors depended on
specific of the external particles (photons in our case) whereas the middle blob
(the phenomenological Pomeron) was universal for all processes with vacuum quantum numbers in the $t$-channel.
This representation holds
for the both BFKL and DL Pomerons. However, there is a technical difference: the impact-factors in the BFKL
concept are
calculated independently of the BFKL Pomeron whereas in DLA they and the middle blob are calculated
universally by constructing
and solving IREEs.

\begin{figure}[h]
  \includegraphics[width=.4\textwidth]{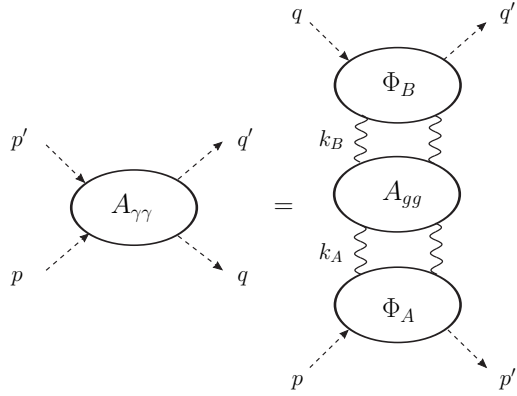}
  \caption{\label{gammaladfig4}Representation of $A_{\gamma \gamma}$ in the form of convolution of impact-factors
  and gluon-gluon scattering amplitude.}
\end{figure}

In both DLA and BFKL  the impact-factors do not have much influence on the high-energy asymptotic behavior of
amplitudes $A_{\gamma \gamma}$. The major role here is played by the middle blob n Eq.~
(\ref{bfklconv}):   the gluon-gluon amplitude $A_{gg}$
is the object fastest growing with $s$. It makes the high-energy asymptotics of $A_{\gamma \gamma}$
be of the Regge form, so its asymptotics is either BFKL or DL Pomeron, depending on the approach.

There is another similarity between the BFKL and DL Pomerons:
The NLO BFKL Pomeron intercepts contains  $\alpha_s$ at an unknown scale. Setting of the scale
was done in Ref.~\cite{kim}.  In contrast,
the DL Pomeron intercept does
not contain $\alpha_s$ at all but contains the IR cut-off $\mu$, which involves a procedure of specifying $\mu$.
Such specifying is universal for singlet and non-singlet Reggeons calculated in DLA (see for
detail Ref.~\cite{egtsum}).

Despite that $\omega_0^{fix~g}$ is close the LO BFKL Pomeron intercept, and
$\omega_0$ is pretty close to  the NLO BFKL Pomeron intercept,
 we do not see any theoretical reason for such similarity and think
that it is just a coincidence. These approaches account for totally different logarithmic contributions: BFKL deals with
single-logarithmic
contributions
and DLA sums DL contributions.

One more difference between BFKL and our approach is that the parameter $s_0$ in the BFKL factor
$(s/s_0)^{\omega_0}$ also requires a special setting procedure which is even more
complicated than setting of the $\alpha_s$ scale (see Ref.~\cite{iv} and references therein) because it includes dealing with the impact-factors
involving many-parton $t$-channel states.
This problem is absent in our approach. For instance, the factor $(Q^2_1Q^2_2)^{1/2}$ in Eq.~(\ref{asam}) is rigorously fixed by
the saddle-point method.

Studying the $s$-behavior of the ratio $R$ of the DL Pomeron to the parent amplitude
allowed us to estimate in Eq.~(\ref{smin}) the applicability region of the
DL Pomeron. A similar estimate for the applicability region of the BFKL Pomeron
has not been done.

The LO BFKL equation\cite{lobfkl} and the approaches based on it sum LL terms $\sim s c^{LO}_n(\alpha_s \ln s)^n$  (see Eq.~\ref{llser}).
The NLO BFKL equation\cite{nlobfkl} and the approaches based on it in addition to the LL terms account for the
sub-leading terms  $\sim s c^{NLO}_n \alpha_s (\alpha_s \ln s)^n$:

\begin{equation}\label{nloser}
A_{NLO} = s \left[1 + \alpha_s \left(c^{LO}_1 \ln s + c^{NLO}_1 \alpha_s \ln s\right) +..
+ \alpha^n_s \left(c^{LO}_n \ln s + c^{NLO}_n \alpha_s \ln s\right)^n +..   \right],
\end{equation}
where $c^{LO}_k$ and $ c^{NLO}_k$ are numerical factors for the leading and sub-leading contributions respectively.
Judging by the involved parameters, the NLO contribution should be much less than the LO one. Indeed,

\begin{equation}\label{lonlo}
s(\alpha_s \ln s)^n \gg s\alpha_s (\alpha_s \ln s)^n.
\end{equation}

The estimate (\ref{lonlo}) predicts that the NLO terms can bring a small impact on the intercept of the
BFKL Pomeron.
However on the contrary,
the NLO contribution  turned out to be of the same order as the LO one. We explain the failure of
the prediction by the fact that
the comparison (\ref{lonlo}) deals with the involved parameters like $s$, $\alpha_s$ but ignores
  the numerical factors $c^{LO}_n$ and $c^{NLO}_n$ because
it is impossible to estimate them
before the calculations have been done. It drives us to conclude that
the by-parameters-considerations like (\ref{lonlo}) are
too rough to predict reliable estimates of numerical characteristics like intercepts.
A similar
situation occur with estimating the impact of the DL contributions~(\ref{dlser}):
if one considers the involved parameters only, the DL terms look  small compared to the LL ones
but as a matter of fact the DL impact on the Pomeron intercept is quite essential.

\section{Summary and outlook}

We have obtained explicit expressions  for the elastic $\gamma \gamma$-scattering amplitudes $A_{\gamma \gamma}$ in DLA  and at the same
time accounted for the running $\alpha_s$ effects. Amplitudes $A_{\gamma \gamma}$ were calculated first in the
collinear kinematics with $t = 0$, see Eqs.~(\ref{am}, \ref{ad}), then expressions~(\ref{mm}, \ref{md}) were obtained for the light-by-light
scattering amplitudes $M_{\gamma \gamma}$ in the forward kinematics with non-zero $t$ and for arbitrary relations between $t$ and the photon virtualities
$Q^2_{1,2}$.  For $A_{\gamma \gamma}$ and $M_{\gamma \gamma}$ we considered separately the cases of deeply-virtual and moderately-virtual
external photons.

Applying the saddle-point method to $A_{\gamma \gamma}$, we obtained its the high-energy asymptotics $A^{as}_{\gamma \gamma}$
 in
Eq.~(\ref{asam}) which is the new, DL  Pomeron. This Pomeron is supercritical like the BFKL Pomeron.
The value of the DL Pomeron intercept depends on accuracy of the calculations, decreasing  with the increase of accuracy, as
shown in Eqs.~(\ref{omega0fixg}-\ref{delta}).
The maximal intercept (\ref{omega0fixg}) corresponds to the roughest approximation where $\alpha_s$ is fixed and quark contributions are
neglected and the minimal intercept is obtained when the both running $\alpha_s$ effects and quark contributions are accounted for.
It drives us to conclude that the further increase of accuracy may diminish intercept of the DL Pomeron  from supercritical values
down to unity, which would agree with the Froissart bound and thereby could prevent violation of Unitarity.

Expressions for the Regge asymptotics are much simpler than for their parent amplitudes and because of that
the asymptotics have often been applied instead of their parent amplitudes to description of high-energy
processes. However it can be done only if the applicability regions for such approximation are known.
To this end we estimated in Eq.~(\ref{smin}) the minimal value of $s$, where the asymptotic $A^{as}_{\gamma \gamma}$ reliably
represent its parent amplitude $A_{\gamma \gamma}$. At $s \leq s_{min}$ one should use
$A_{\gamma \gamma}$ instead of $A^{as}_{\gamma \gamma}$.

Our results do not mean that the DL Pomeron is supposed to substitute BFKL Pomeron. On the contrary, it would be challenging to study
interference between contributions of the both Pomerons to high-energy reactions. 
Finally, we would like to notice that  
the explicit expressions for $A_{\gamma \gamma}$ obtained in our paper should be used at available energies 
rather than their asymptotics. 

\section{Acknowledgement}

We are grateful to V.T.~Kim, A.V.~Kotikov, D.A.~Ross,  W.~Schafer and especially to D.Yu.~Ivanov  for interesting discussions.

\end{document}